\documentclass[fleqn,twoside]{article}
\usepackage{espcrc2}
\usepackage{subfigure}
\usepackage{graphicx}
\usepackage[figuresright]{rotating}

\newcommand{\mmu}{\mbox{\ensuremath{m_{\mu}}}}

\newcommand{\amu}[1][]{\ensuremath{a_{\mu^{#1}}}}
\newcommand{\gm}{\ensuremath{(g-2)}}
\newcommand{\wa}{\mbox{\ensuremath{\omega_a}}}
\renewcommand{\wp}{\mbox{\ensuremath{\omega_p}}}

\title{Measurement of the muon anomaly to high and even higher precision}

\author{David W. Hertzog\address{Department of Physics \\
University of Illinois at Urbana-Champaign, Urbana, IL 61801
USA}\thanks{Representing the E821~\cite{E821} and the
E969~\cite{E969} Collaborations.}}

\begin{document}
\begin{abstract}
Our recent series of measurements at Brookhaven National
Laboratory determined the muon anomalous magnetic moment \amu\ to
a precision of 0.5~ppm. The final result---representing the
average of five running periods using both positive and negative
muons---is $\amu ^\pm = 11\,659\,208(6) \times 10^{-10}$. It lies
2.7 standard deviations above the standard model expectation,
which is based on updates given at this Workshop.  Importantly,
only the $e^{+}e^{-}$ annihilation and new KLOE radiative return
data are used for the hadronic vacuum polarization input. Because
the systematic limit has not been reached in the experiment, a new
effort has been proposed and approved with the highest scientific
priority at Brookhaven. The goal is an experimental uncertainty of
0.2~ppm, a 2.5-fold reduction in the overall experimental
uncertainty. To do so will require a suite of upgrades and several
qualitative changes in the philosophy of how the measurement is
carried out. I discuss the old and new experiments with a
particular emphasis on the technical matters that require change
for the future.

\end{abstract}

\maketitle

\section{Introduction}
The muon anomalous magnetic moment \amu\ is one of the most
precisely measured and calculated quantities in physics. The
standard model (SM) theoretical expectation and the recent
experimental measurements have both achieved sub-ppm precision,
enabling a sensitive comparison having new-physics ramifications.
A significant deviation implies ``missing'' physics in the model,
which is expected to occur in the most popular SM extensions such
as SUSY. Here we discuss the status of the most recent experiment
and the recently approved new experiment aimed at pushing the
precision frontier further by a factor of approximately 2.5.  We
give a brief review of the current standard model expectation.

The E821 Brookhaven National Laboratory (BNL)
experiment~\cite{E821} is the fourth in a sequence of \amu\
measurements. The rich history includes three ingenious efforts at
CERN, the final one~\cite{Bailey:1977mm} achieving a precision of
7.3~ppm. Our BNL Collaboration took data from 1997 through 2001.
The analysis is complete and all results have been
reported~\cite{Carey:1999dd,Brown:2000sj,Brown:2001mg,Bennett:2002jb,Bennett:2004xx}
and a comprehensive review has been
published~\cite{HertzogMorse2004}.  With nearly equally precise
\amu\ determinations from positive and negative muon samples---the
results are perfectly compatible---the CPT-combined final result
is $\amu ^\pm = 11\,659\,208(6) \times 10^{-10}$. The statistical
and systematic uncertainties are combined in quadrature and the
0.5~ppm precision represents a 15-fold improvement compared to
CERN-III. The uncertainty is dominated by statistics so it is
natural to contemplate an upgrade program aimed at increasing the
data accumulation rate while maintaining or reducing the
systematic uncertainties. To this end, the E969
Collaboration---formed largely from the present group plus key new
institutions---has proposed and has been approved for new running.
Several significant conceptual changes, which will increase the
data rate fivefold or more and reduce systematic uncertainties
further, are required.

The standard model expectation for the muon anomaly is based on
QED, hadronic and weak loop contributions.  Davier and Marciano
have recently published a review of the theory~\cite{davmar} and
at this Workshop, numerous speakers provided technical updates to
specific contributions. I refer the reader to their expert reports
in these proceedings and summarize the current standard model
expectation briefly in Table~\ref{tab:theory}.  Only the
$e^{+}e^{-}$ annihilation data are used, together with the new
radiative return results from KLOE~\cite{KLOE}, as the low-energy
input to the 1st-order hadronic vacuum polarization contribution.
The hadronic light-by-light scattering contribution has been
addressed by many authors.  I use the recommendation in
Ref.~\cite{davmar}, which attempts to find a central value with an
expanded uncertainty to accommodate different individual
evaluations.

\begin{table*}[hbt]
\caption{Standard model theory and experiment
comparison.\label{tab:theory}}
\begin{tabular}{lrrcl}
\hline
Contribution & Value  & Error & Reference & Comment \\
             & $\times 10^{10}$ & $\times 10^{10}$ & & \\
\hline
QED & 11\,658\,471.94 & 0.14 & \cite{kinqed} & 4 loops; 5th estimated \\
Hadronic vacuum polarization~~~~~~ & 693.4 & 6.4 & \cite{daviertau04} & $e^{+}e^{-}$  + KLOE \\
Hadronic light by light &   12.0 & 3.5 & \cite{hlbl} & val. from Ref.~\cite{davmar} \\
Hadronic, other 2nd order & -10.0 & 0.6 & \cite{dehz2} & alt.: $-9.8\pm0.1$\\
Weak & 15.4 & 0.22 & \cite{davmar} & 2 loops \\
\hline Total theory & 11\,659\,182.7 & 7.3 & -- & -- \\
Experiment & 11\,659\,208 & 6 & \cite{Bennett:2004xx} & -- \\
Expt. - Thy. & 25.3 & 9.4 & -- & 2.7 standard deviations\\ \hline
\end{tabular}
\end{table*}

\section{The "Old"  BNL Experiment: E821}
\begin{figure}[hbt]
\includegraphics*[width=0.95\columnwidth]{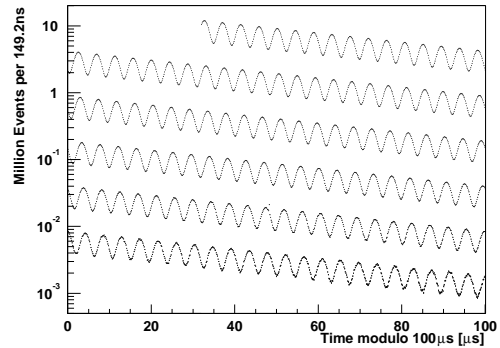} \caption{Distribution
of counts versus time for the 3.6 billion decays in the 2001
negative muon data-taking period. \label{fig:wiggles}}
\end{figure}

The central element of the BNL experiment is the muon storage
ring~\cite{ring}, which has a highly uniform 1.45~T magnetic
field, a 7.1~m radius, and vertical containment of the muons by
electrostatic quadrupoles~\cite{quads}. At the magic momentum of
3.094~GeV/$c$ the relativistic gamma is 29.3, the dilated muon
lifetime is 64.4~$\mu$s, and the decay electrons\footnote{By
convention, we discuss negative muons and their decay electrons
throughout this paper.} have an upper lab-frame energy of
approximately 3.1~GeV. The basic idea of the measurement is to
inject and store a bunch of polarized muons into the ring, contain
them sufficiently well while they make many revolutions, and
record their decay time by detecting the emitted electrons that
curl inward and strike electromagnetic calorimeters. If $g=2$, the
initial muon spin direction would remain aligned with its momentum
and the rate of decay electrons in the detectors would follow a
smoothly falling exponential. For $g > 2$, the muon spin precesses
at a rate faster than the muon rotation rate; this additional
rotation---the difference between the spin rotation and the
cyclotron frequency---is directly proportional to \amu. The
average muon spin direction is revealed because parity violation
in the weak muon decay correlates the electron energy (on average)
with the muon spin direction. The rate of detected electrons
having an energy greater than a set threshold is an exponential
(as above), but modulated at the anomalous precession frequency,
see Fig.~\ref{fig:wiggles}. The key to the experiment is to
determine this frequency to high precision and to measure the
average magnetic field to equal or better precision.

The anomalous precession frequency \wa\ depends on the cyclotron
frequency and the spin rotation.  In the presence of a magnetic
and electric field having $\vec {B} \cdot \vec {P} = \vec {E}
\cdot \vec {P} = 0$, \wa\ is described by
\begin{eqnarray}
\label{eq:spin-precess}
 \vec{\omega _{a}}  & \equiv & \vec{\omega}_a - \vec{\omega}_c
  \\
& = & \frac{e}{\mmu c} \left[\amu\vec{B} - \left(\amu -\frac
 {1}{\gamma^2-1}\right)(\vec{\beta} \times \vec{E})\right]\nonumber.
\end{eqnarray}

The term in parentheses multiplying $\vec{\beta} \times \vec{E}$
vanishes at $\gamma = 29.3$ and the electrostatic focussing does
not affect the spin (except for a correction necessary to account
for the finite momentum range  $\Delta P/P \approx \pm0.14\%$
around the magic momentum). Equation~\ref{eq:spin-precess} can be
rearranged to isolate \amu, giving $\wa /B$ multiplied by physical
quantities that are known to high precision. The experiment is
described below with emphasis on various specific details that
will be upgraded for the new E969 effort.

Polarized muons are obtained from the decay $\pi^- \rightarrow
\mu^{-}\bar{\nu}_{\mu}$.  In the massless neutrino limit, helicity
conservation implies that the outgoing anti-neutrino is
right-handed; its spin is in the direction of its motion. The
negative muon---emitted in the opposite direction---is polarized
along its motion.  In the CERN-III effort, a 3.1~GeV/$c$ pion beam
was injected into a storage ring. Decay muons, born during the
first turn of the beam in the ring, could fall (with very low
probability, $\approx~100$~ppm) onto stable orbits within the
storage ring aperture. The captured muons came from the nearly
forward-emitted muons, which had the right momentum but also a
small transverse decay ``kick'' that placed them in the storage
ring acceptance. The majority of the pions struck detectors or the
storage ring yoke causing an enormous prompt hadronic ``flash'' in
the detectors, which was followed by a slowly diminishing
background from the thermalization and capture of neutrons.

\begin{figure}[hbt]
  \begin{center}
    \includegraphics*[width=\columnwidth]{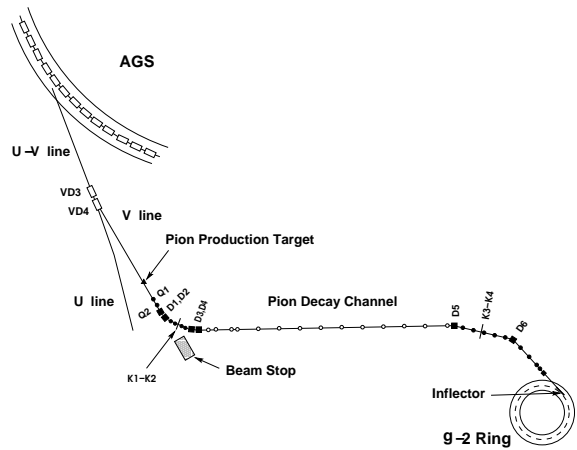}
    \caption{Plan view of the pion/muon beamline. \label{fig:beamline}}
  \end{center}
\end{figure}

E821 improved on this method of muon loading significantly by
directly injecting an enriched muon beam into the storage ring and
then using a pulsed magnetic kicker~\cite{kicker} to deflect the
muons into the ring acceptance. A pion-muon beamline upstream of
the ring was developed to serve three tasks: 1) a pion creation
and capture section; 2) a pion-to-muon decay section; and 3) a
final muon selection at $P_{magic} = 3.094~$GeV/$c$.
Figure~\ref{fig:beamline} is a schematic of the beamline and
storage ring. The pion capture and muon selection sections are
tuned to slightly different momenta. Typical operating conditions
set the pion momentum $1.7\%$ greater than $P_{magic}$. Forward
decay muons, but slightly off axis from zero degrees, are captured
in the decay channel. A final bend in the beamline is made by
setting dipoles D5 and D6 to transport particles having momentum
$P_{magic}$.  This excludes a large fraction of the pions, because
their average momentum is higher.  Slits K3/K4 adjust the momentum
bite that is passed into the storage ring. The ratio of pions to
muons entering the ring through the superconducting inflector is
roughly 1:1 at nominal settings.  The ratio can be adjusted:
higher muon rates are obtained at the expense of greater pion
contamination; lower rates are obtained for lower pion
contamination. The employed operating conditions represented an
appropriate compromise.

Direct muon injection resulted in an approximately 10-fold
increase in the effective stored muon rate and a reduction of the
hadronic flash by a factor of 50. But the hadronic flash is not
eliminated by this technique.  The pions that accompany the muons
into the ring still cause significant neutron production and
create a prompt flash at injection. Because of this, each of the
24 electromagnetic calorimeters is gated off during injection and
turned back on 5 to 30~$\mu$s after injection; the turn-on time
depends on the location in the ring. When the physics fits start,
typically 30~$\mu$s after injection, the photomultiplier tubes
(PMTs) are still affected by a slowly decaying background, which
appears as a time-dependent pedestal in the recorded digitized
pulses. The electron rate in each detector at this time is
typically several MHz.

Parity violation in the muon weak decay $\mu^- \rightarrow e^-
\bar{\nu}_e \nu_{\mu}$ allows the spin direction to be measured as
a function of time (on average).  The angular distribution of
emitted electrons from polarized muons at rest is $dn/d\Omega = 1
- a(E)\cos\theta$ where $\theta$ is between the spin and momentum
directions and the asymmetry $a$ depends on electron energy $E$.
Expressions for $n$ and $a$ are:
\begin{eqnarray}
n(y) & = & y^{2}(3-2y) \nonumber \\
a(y) & = & (2y-1)/(3-2y),
\end{eqnarray}
with $y = P_{e}/P_{e,max}$. The asymmetry is negative for
low-energy electrons and rises to $a = +1$ when $y = 1$. The
higher-energy electrons have the strongest correlation to the muon
spin. In the lab frame, it is most probable to detect a
high-energy electron when the muon spin is pointing opposite to
the direction of muon momentum and least likely when the spin is
aligned with the muon momentum.  The rate of detected electrons
above energy threshold $E_{th}$ is
\begin{equation}
 \frac{dN(t;E_{th})}{dt} = N_{0} e^{-t/\gamma \tau_{\mu} } \left[1 +
 A\cos(\wa t + \phi\right)], \label{eq:fivepar}
\end{equation}
where $A$ is the effective integrated asymmetry with $A \approx
0.4$ for $E_{th} = 1.9$~GeV. Equation~\ref{eq:fivepar} is the
simplest expression that describes the data. In practice, it can
only be used to obtain an estimate of parameters because
perturbations to the spectrum are caused from beam dynamics or
detector and electronic effects. For example, the natural betatron
oscillations of the stored muon ensemble beat with the cyclotron
frequency causing an observable modulation in the normalization
term $N$ above because of a modulation in the effective detector
acceptance. In a more subtle manner, the modulation extends to the
asymmetry and phase terms. Equation~\ref{eq:fivepar} can be easily
modified to account for these modulations. Similarly, gain (or
energy-scale) stability, pileup, and muon loss terms require small
corrections or fit-function modifications in order to obtain a
good fit. But these are well-understood procedures having
carefully understood implications on the extracted precession
frequency \wa.  In general, they do not couple strongly to \wa,
but they do affect the goodness-of-fit criterion.

\begin{figure*}[hbt]
\begin{center}
\includegraphics*[width=\textwidth]{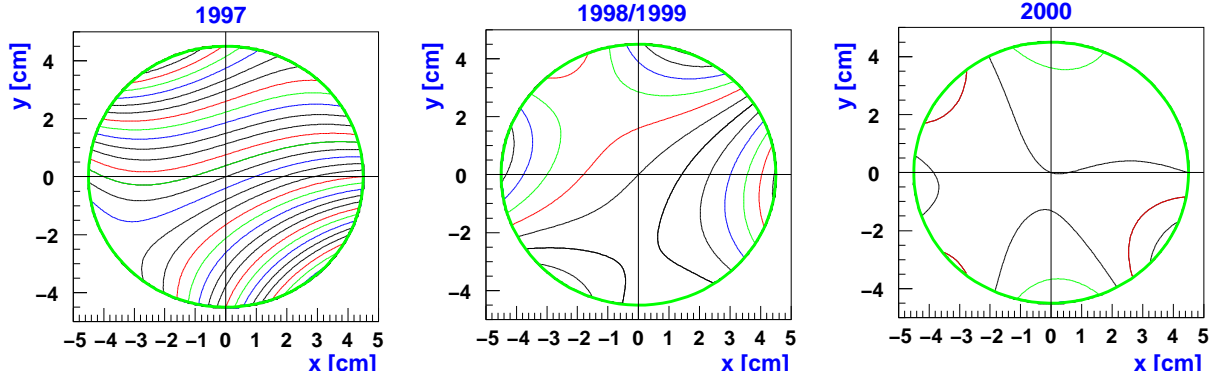}
\caption{Sequence of improvements to the magnetic field profiles
since the original commissioning of the storage ring. The contours
are averaged over azimuth and interpolated using a multipole
expansion. The circle indicates the storage aperture. The contour
lines are separated by 1~ppm deviations from the central average.
From left to right, these maps represent the field variations for
the 1997, 1998/1999 and 2000 time periods, respectively.  The 2001
map, with reversed polarity field, was slightly more uniform than
the 2000 map. \label{fig:fieldmaps}}
\end{center}
\end{figure*}

Equation \ref{eq:spin-precess} also contains the term $B$, which
implies knowing the magnitude of the magnetic field within the
storage volume as a function of time and weighting the data with
the average field.  The best way to make this otherwise daunting
task manageable is to shim the field to high uniformity.  The
uniformity is well illustrated by the three contour maps shown in
Fig.~\ref{fig:fieldmaps}.  Each figure represents an azimuthially
averaged magnetic field from the beginning of commissioning in
1997, to the first muon-injection runs in 1998/1999, and later to
the high-statistics run in 2000. The contours in these figures
represent 1~ppm deviations from the central average field value.
The final 2001 field, with reversed polarity, was equally good to
the 2000 map.  The field is very uniform, not only for the present
experiment, but also for the future.

All field measurements rely on proton NMR using a system of probes
that obtains the absolute field value, the relative field versus
time, and the relative field inside the storage volume.  This
system~\cite{nmr} uses a standard reference probe, a ``plunging''
probe that is inserted into the vacuum chambers to transfer the
calibration, an NMR ``trolley'' to map the field inside the
storage ring volume, and 378 fixed probes located in the vacuum
chamber walls to monitor the field versus time. The magnetic field
is measured in units of the free proton precession frequency,
$\omega_p$. Muonium measurements determine the ratio $\lambda$ of
muon-to-proton magnetic moments~\cite{liu} and, in practice, \amu\
is determined from the expression

\begin{equation}
\amu = \frac{R}{\lambda - R},  \label{eq:amuR}
\end{equation}
where $R = \wa/\wp$, a ratio determined in our experiment.

\begin{figure*}
\begin{center}
\label{tab:aMuSummary}
\includegraphics*[width=.75\textwidth]{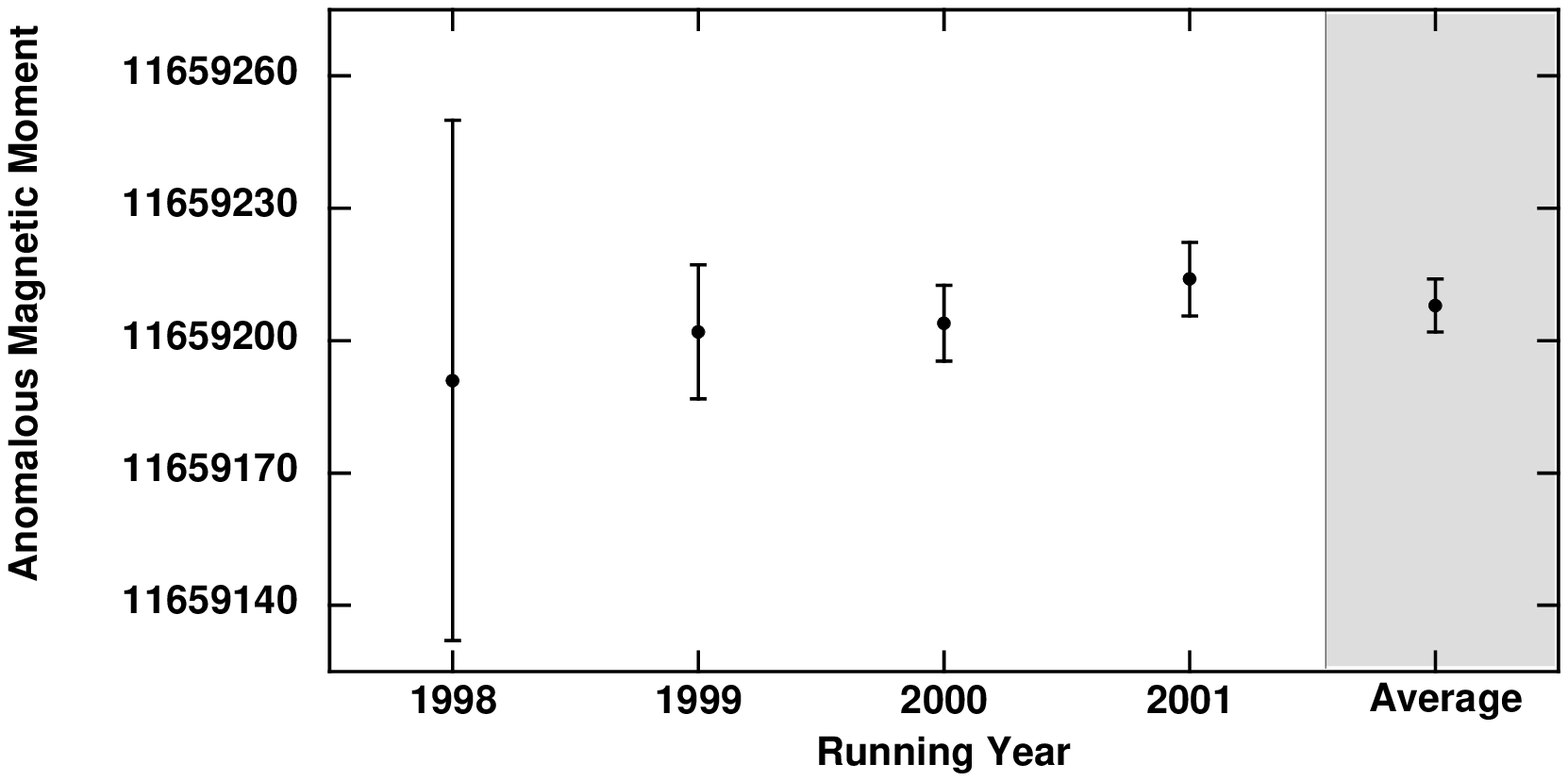}
\caption{Compilation of results from E821 from 1998 - 2001 running
periods, together with the average. \label{fig:summary}}
\end{center}
\end{figure*}

In each running year, multiple analyses of both the precession
frequency and the magnetic field data were carried out
independently.  Reports to the collaboration were always made with
a secret offset so that no one person could compute \amu. After
all results were final and all systematic uncertainties
established, the offsets were removed and \amu\ was computed.
Figure~\ref{fig:summary} illustrates the consistency of the
results from each running year beginning with 1998 and
Table~\ref{tab:aMuSummary} gives all of the numeric results.

\begin{table*}[hbt]
\caption{Summary of E821 Results}
\begin{tabular}{cclcc}
\hline
Year & Polarity & $a_\mu \times 10^{10}$ & ~~~Precision [ppm]~~~ & Reference \\
\hline
1997~~~~ & ~~~~$\mu^+$~~~~ &   11\,659\,251(150) & 13 & ~~\cite{Carey:1999dd}~~ \\
1998 & $\mu^+$ &   11\,659\,191(59) & 5 & \cite{Brown:2000sj} \\
1999 & $\mu^+$ &   11\,659\,202(15) & 1.3 & \cite{Brown:2001mg} \\
2000 & $\mu^+$ & 11\,659\,204(9) & 0.7 & \cite{Bennett:2002jb} \\
2001 & $\mu^-$ & 11\,659\,214(9) & 0.7 & \cite{Bennett:2004xx} \\
\hline
Average & & 11\,659\,208(6) & 0.5 & \\
\hline
\end{tabular}
\end{table*}

A precision measurement depends on control of the systematic
uncertainties.  Table~\ref{tab:syster} lists all the principle
uncertainties for the last three high-statistics running periods.
The categories are separated by the independent \wa\ and \wp\
analyses. The trend to reduce the uncertainties in each year is
evident, except for the CBO effect, which was only fully
understood prior to the final 2001 run so that running parameters
could be adjusted to minimize its effect.  The goal for the new
experiment is to limit the field and the precession systematic
uncertainties each to 0.1~ppm.  New data taking operation modes of
the storage ring are being planned to reduce some of the
beam-dynamics related uncertainties such as CBO and muon loss.

\begin{table*}[hbt]
\caption{Systematic Errors from the 1999, 2000 and 2001 data
sets~\cite{Brown:2001mg,Bennett:2002jb,Bennett:2004xx}.
$^{\dag}$Higher multipoles, the trolley frequency, temperature,
and voltage response, eddy currents from the kickers, and
time-varying stray fields \hfill\break
 $\ddag$In 2001  AGS background, timing shifts, E field and
vertical oscillations, beam debunching/randomization, binning and
fitting procedure together equaled 0.11 ppm } \label{tb:syster}
\begin{tabular}{|l|c|c|c||l|c|c|c|}
\hline $\sigma_{\rm syst}$  $\omega_p$  &1999 & 2000 & 2001 &
$\sigma_{\rm syst}$ $\omega_a$  &1999  &  2000 &2001 \\
  & (ppm) & (ppm)  &  (ppm) & & (ppm) & (ppm) & (ppm) \\
\hline Inflector Fringe Field & 0.20 &  -  & - &
Pile-Up & 0.13 & 0.13  & 0.08    \\
Calib. of trolley probes & 0.20 & 0.15  & 0.09   &
AGS background & 0.10 & 0.01  &  $\ddag$   \\
Tracking $B$ with time & 0.15 & 0.10 & 0.07  &
Lost Muons & 0.10 & 0.10 & 0.09  \\
Measurement of $B_0$ & 0.10 & 0.10 &0.05  &
Timing Shifts & 0.10 & 0.02  & $\ddag$  \\
$\mu$-distribution & 0.12 & 0.03 & 0.03  &
E-field/pitch & 0.08 & 0.03  & $\ddag$  \\
Absolute calibration  & 0.05 & 0.05 & 0.05   &
Fitting/Binning & 0.07 & 0.06  & $\ddag$ \\
Others$^{\dag}$  & 0.15 & 0.10 & 0.07  &
CBO & 0.05 & 0.21 & 0.07  \\
  & & &  &
Beam debunching & 0.04 &0.04 & $\ddag$ \\
  & & &  &
Gain Changes & 0.02 & 0.13 & 0.12 \\
\hline Total for $\omega_p$ & {0.4} & {0.24} & 0.17 &
Total for $\omega_a$ &{0.3} &{0.31}  &  0.21  \\
\hline
\end{tabular}

\end{table*}

\section{The "New" BNL Experiment: E969}
In September, 2004, the BNL Program Advisory Committee gave the
proposal {\em A (g-2)$_{\mu}$ Experiment to $\pm0.2$~ppm
Precision} its highest scientific endorsement, thus officially
launching experiment E969~\cite{E969}. The central idea is to use
the existing BNL storage ring largely as is but to develop a means
to increase the stored muon rate, reduce the background, and to
maintain or reduce individual systematic uncertainties.

\subsection{Backward decay and beamline modifications}
When the pion capture section of the beamline is tuned 0.5\% above
the muon magic momentum, the highest muon flux is achieved. As
noted, this small momentum difference passes too many pions into
the storage ring, creating background. If the initial pion
momentum is set to 5.32~GeV/$c$, then the {\em backward} decay
muons are produced at 3.094~GeV/$c$ as desired. The mismatch in
momentum between the pion capture section and the muon momentum
selection section is so great that no pions can pass through the
K3/K4 slits shown in Fig.~\ref{fig:beamline}.  This implies that
the entire hadronic flash will be eliminated--a key improvement
allowing the detectors to be gated on much earlier after injection
and permitting the physics fits to start shortly thereafter.
Simulations show that the muons are still highly polarized
(albeit, the direction is reversed) and the flux is at or higher
than the level obtained in the current forward-decay scheme.  To
create the backward-decay beam, several new front-end magnetic
elements must be built because not all of the present magnets can
be ramped up by the momentum ratio factor 5.32/3.15.  New magnet
designs are being pursued and new front-end tunes are being
developed by our team.

Beam transport studies of the pion-to-muon decay section indicate
that improved transmission can be obtained by doubling or tripling
the number of quadrupole elements in the line. Currently, the 80~m
long FODO section has 20 magnets but ample room exists for up to
four times as many.  For both the forward-decay or the new
backward-decay kinematics, the decay muons are captured with a
wide angular divergence compared to the pion source. The increased
quadrupole density maintains the lateral extent of the secondary
muon beam within the magnet bore, thus avoiding losses that
presently exist.  A gain in rate by a factor of 2 is expected from
this straight-forward improvement.

\begin{figure*}[hbt]
\begin{center}
\subfigure[Closed End
]{\includegraphics[width=0.45\textwidth]{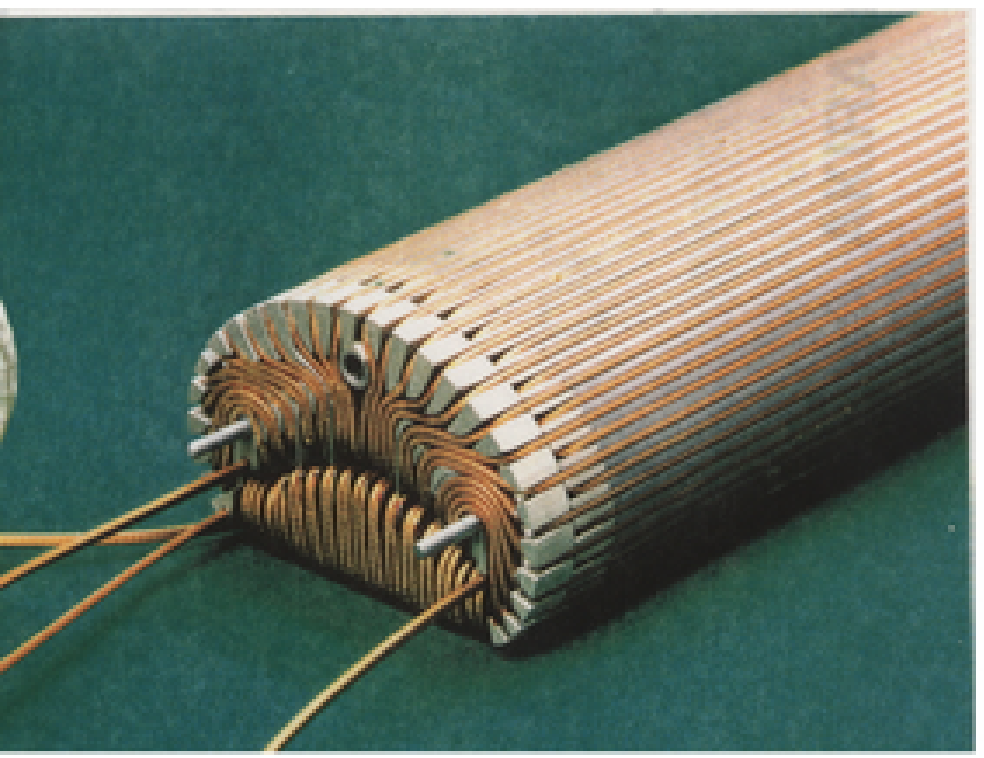}}
\subfigure[Open
End]{\includegraphics[width=0.45\textwidth]{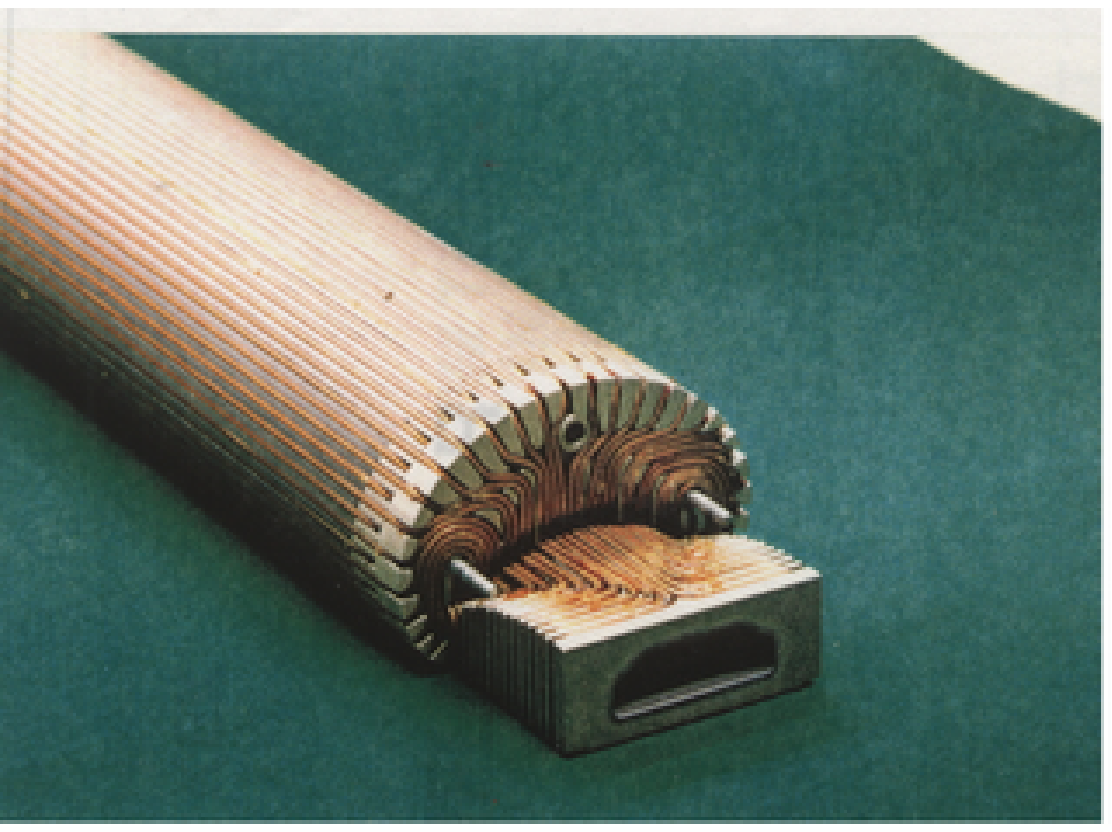}}
\end{center}
\caption{Photos of the closed- and open-end inflector prototype.
The inflector is turned on its side. The opening, evident in (b),
has an 18~mm width and 57~mm height.} \label{fig:inflector}
\end{figure*}

The muon beam enters the storage ring through a superconducting
inflector~\cite{inflector}.  The inflector has coils that
essentially block both the entrance and exit openings.  Muons lose
energy and scatter as they pass through these coils and
approximately half of those muons are lost.  A closed-ended
inflector is shown in the photograph in Fig.~\ref{fig:inflector}a.
The full inflector was built with this coil design because it
represented a practical choice for reliability reasons at the time
of production. However, the opposite end of the inflector
prototype had an open end (see Fig.~\ref{fig:inflector}b), which
performed as well as the closed end in magnetic tests. We intend
to build a new full-scale inflector following this design having
both ends open to regain the factor of 2 in lost muons.

\subsection{Electron detection and data collection}
E821 has 24 lead-scintillating fiber electromagnetic
calorimeters~\cite{detectors} to detect electron energy and time
of arrival.  Each 14~cm high by 23~cm wide calorimeter is
12.5~$X_0$ deep (15~cm) and the fiber direction is radial.  Four
lightguides transport the light to individual PMTs, located below
the storage ring midplane to avoid being directly struck during
the hadronic flash.  Waveform digitizers sample the summed PMT
signals every 2.5~ns.  A sequence of these samples is the record
of a given event.  These records are scanned for electron pulses
and fit to determine energy and time. Pileup occurs when two
pulses strike the detector within the resolving time of the system
(typically $<5~$ns).  Pileup events can be removed, on average, by
a procedure that uses the leading and trailing samples around a
triggered electron pulse, however, a systematic uncertainty
remains from incomplete pileup subtraction and from pulses too low
in energy to be accounted for properly. Because the rates will
increase by approximately fivefold in E969, pileup will dominate
the systematic uncertainties if left unimproved. A new calorimeter
will be built and a new parallel data-taking method is being
developed to address the anticipated higher rates. Additionally, a
more finely sampling waveform digitizer will be used and the
fivefold vertical segmentation of the existing front scintillator
detector hodoscopes will be increased by at least a factor of 2.

\begin{figure}[hbt]
\begin{center}
\includegraphics*[width=.7\columnwidth]{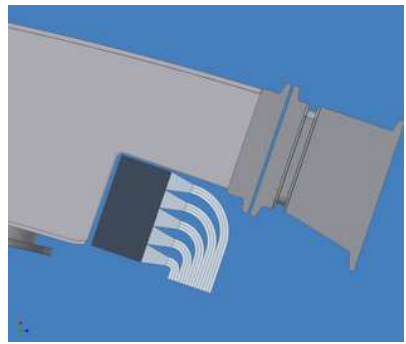}
\caption{Plan view of new calorimeter in the region of the bellows
between vacuum chambers where the available space for lightguides
is quite restricted. \label{fig:calo-vacuum-bellows}}
\end{center}
\end{figure}

The new calorimeter will be dense and fast and it will be
segmented transversely with respect to the incoming electrons.
Readout of individual segments will occur on the downstream side
where space is severely limited by existing vacuum chamber
flanges. {\tt GEANT} studies show that pileup events can be
recognized 4 out of 5 times---which is the goal to reduce this
systematic uncertainty---if the Moliere radius is restricted to
less than 1.7~cm. We have developed a preliminary design using
tungsten and scintillating fiber ribbons that meets these
requirements. The calorimeter will be subdivided into twenty $4
\times 4$~cm modules, each having a length of 11~cm ($15 X_0$).
The individually read-out modules are stacked in a 4 high by 5
wide array. Laminated thin acrylic sheets are bundled and bent at
90~degrees toward the inside of the ring to pipe the light to
20~PMTs. A sketch is shown in Fig.~\ref{fig:calo-vacuum-bellows},
where the geometrical constraints are evident.

\subsection{The $Q$ method of data accumulation}
The traditional ``$T$'' method of data selection in E821 involves
accepting electrons only when their energy exceeds a fixed
threshold. Such events are sorted into histograms of events versus
time with weight 1 for each event.  In contrast, the ``$Q$''
method does not rely on the separate identification or isolation
of events. It simply histograms the energy striking the detectors
versus time.  All electrons that shower in the calorimeters are
included, but their decay time is intrinsically weighted by their
energy. The net asymmetry is roughly halved compared to the $T$
method but it is compensated for by the use of all events.

Full {\tt GEANT} simulations were used to compare the methods,
finding that the $Q$~method is statistically weaker by about
9~percent. However, it requires no pileup correction. In fact, for
the $Q$ method, the detectors do not need to be segmented at all
because only the integrated energy deposition is of interest. We
plan to employ both data-taking methods in E969.

\subsection{Magnetic field determination}
The magnetic field will be measured to approximately 0.11\,ppm
using the same technique and apparatus as was used in E821. The
uncertainty in the 2001 run was 0.17~ppm and this will be
improved.  The proposal~\cite{E969} details the work plan:
\begin{itemize}
\item{Measurement of the field change from kicker
eddy currents \emph{in situ}~\cite{kicker},}
\item{Extensive measurements with the magnetic field trolley,
aiming in particular to better resolve the position of the active
NMR volumes inside the trolley shell and to map out the response
functions to the level where \emph{corrections} can be applied,
rather than \emph{limits} be set,}
\item{More frequent measurements of the magnetic field in the
storage ring during beam periods,}
\item{Repair and retuning of a number of fixed NMR probes to improve
 the sampling of the storage ring,}
\item{Analysis refinements to reduce
trolley position uncertainties.}
\end{itemize}

\section{Summary}
Our E821 experiment is complete.  The result at 0.5~ppm precision
is statistics limited. This brief overview was aimed at
illustrating a few of the technical issues that must be addressed
to continue the experiment at higher rates and the same or reduced
systematic uncertainties.  With the full complement of plans
realized, we project that an overall uncertainty of 0.2~ppm can be
obtained in one run of roughly 20 weeks, which would follow
several years of construction and a short commissioning run. The
Collaboration is beginning to do the simulation work necessary and
a formal Technical Design Report will be written for proper
costing and time projections.

If the past history of $\gm_{\mu}$ experiments can be used as a
guide, then the next generation effort will certainly reveal
exciting physics results.  At this Workshop, we have learned of
the extensive worldwide effort to establish the standard model
\amu\ expectation with high precision.  New radiative return data
from KLOE~\cite{KLOE} supports the hadronic vacuum polarization
determination, which is largely based on the precise CMD-II
$e^{+}e^{-}$ annihilation measurements. Additional data are
expected from BaBar and Belle. Meanwhile, the discrepancy between
$e^{+}e^{-}$-based and hadronic tau-based spectral functions
remains.  Until this is resolved, Davier and H\"ocker suggest
using the $e^{+}e^{-}$-based results, which is reflected in
Table~\ref{tab:theory}. Many contributions to this Workshop have
helped to clarify the issues related to the hadronic contributions
to the standard model expectation. Projecting to the future, we
believe that the theoretical uncertainty can be reduced to
$\approx~0.4$~ppm. Combined with an experimental uncertainty of
0.2~ppm, this will yield an improved sensitivity to new physics by
a factor of 2. The current $2.7\sigma$ difference between
experiment and theory is tantalizing, but not definitive. An
additional resolution by a factor of 2 is therefore very exciting
to contemplate and we all eagerly anticipate this next step in
precision.

\section{Acknowledgments}
The author expresses gratitude to the Guggenheim Foundation for
partial support. The \gm~ experiment is supported in part by the
U.S. Department of Energy, the U.S. National Science Foundation,
the German Bundesminister f\"{u}r Bildung und Forschung, the
Russian Ministry of Science, and the US-Japan Agreement in High
Energy Physics.


\begin{thebibliography}{00}
\bibitem{E821} Muon E821 Collaboration:
G.W. Bennett, B. Bousquet, H.N. Brown, G. Bunce, R.M. Carey, P.
Cushman, G.T. Danby, P.T. Debevec, M. Deile, H. Deng, W. Deninger,
S.K. Dhawan, V.P. Druzhinin, L. Duong, E. Efstathiadis, F.J.M.
Farley, G.V. Fedotovich, S. Giron, F.E. Gray, D. Grigoriev, M.
Grosse-Perdekamp, A. Grossmann, M.F. Hare, D.W. Hertzog, X. Huang,
V.W. Hughes\, M. Iwasaki, K. Jungmann, D. Kawall, M. Kawamura,
B.I. Khazin, J. Kindem, F. Krienen, I. Kronkvist, A. Lam, R.
Larsen, Y.Y. Lee, I. Logashenko, R. McNabb, W. Meng, J. Mi, J.P.
Miller, Y. Mizumachi, W.M. Morse, D. Nikas, C.J.G. Onderwater, Y.
Orlov, C.S. \"{O}zben, J.M. Paley, Q. Peng, C.C. Polly, J. Pretz,
R. Prigl, G. zu Putlitz, T. Qian, S.I. Redin, O. Rind, B.L.
Roberts, N. Ryskulov, S. Sedykh, Y.K. Semertzidis, P. Shagin,
Yu.M. Shatunov, E.P. Sichtermann, E. Solodov, M. Sossong, A.
Steinmetz L.R. Sulak, C. Timmermans, A. Trofimov, D. Urner, P. von
Walter, D. Warburton, D. Winn, A. Yamamoto and D. Zimmerman.


\bibitem{Bailey:1977mm} J.~Bailey et al., Nucl. Phys. {\bf B150}, 1 (1979).

\bibitem{Carey:1999dd} The $g-2$ Collaboration:
R.M.~Carey et al., Phys. Rev. Lett. {\bf 82}, 1132 (1999).
\bibitem{Brown:2000sj} The $g-2$ Collaboration:
H.N. Brown et al., Phys. Rev. {\bf D62}, 091101 (2000).
\bibitem{Brown:2001mg} The $g-2$ Collaboration:
H.N.~Brown et al., Phys. Rev. Lett. {\bf 86}, 2227 (2001).
\bibitem{Bennett:2002jb}The $g-2$ Collaboration:
G.W. Bennett et al., Phys. Rev. Lett. 89 (2002) 101804;
Erratum-ibid. 89 (2002) 129903.
\bibitem{Bennett:2004xx}The $g-2$ Collaboration:
G.W. Bennett et al., Phys. Rev. Lett. 92 (2004) 161802.

\bibitem{HertzogMorse2004} D.W. Hertzog and W.M. Morse,
Annu. Rev. Nucl. Part. Sci. {\bf 54} 141 (2004).

\bibitem{davmar} M. Davier and W. Marciano,
Annu. Rev. Nucl. Part. Sci. {\bf 54} 115 (2004).
\bibitem{KLOE}A. Aloisio, et al., (KLOE Collaboration)
hep-ex/0407048, July 2004, and Phys. Lett. {\bf B}, in press.
\bibitem{kinqed} T. Kinoshita and M. Nio, Phys.Rev.Lett.90:021803,2003,
and T. Kinoshita and M. Nio, Phys. Rev. {\bf D70}, 113001 (2004).

\bibitem{daviertau04} M. Davier, this Workshop,
including the new KLOE radiative return data, but not the hadronic
tau decay results. K. Hagiwara presented a similar number at this
Workshop but did not yet include the KLOE results, see K.
Hagiwara, A.D. Martin, D. Nomura, and T. Teubner, Phys. Lett. {\bf
B557}, 69 (2003), and Phys. Rev. {\bf D69} 093003 (2004).

\bibitem{dehz2} M. Davier, S. Eidelman, A.~H\"ocker, and Z. Zhang,
Eur. Phys. J. {\bf C 31}, 503 (2003).

\bibitem{hlbl} J. Bijnens, E. Pallante
and J. Prades, Nucl. Phys. {\bf B474}, 379 (1996) and   Nucl.
Phys. {\bf B626}, 410 (2002); M. Hayakawa and T. Kinoshita, Phys.
Rev. {\bf D57}, {465}{(1998)} and hep-ph/0112102 (2002); M.
Knecht, A. Nyffeler, Phys. Rev. {\bf D65}, 073034 (2002); M.
Knecht, A. Nyffeler, M. Perrottet, E. De Rafael, Phys. Rev. Lett.
{\bf 88}, 071802 (2002); I. Blokland, A. Czarnecki and K.
Melinkov, Phys. Rev. Lett. {\bf 88}, 071803 (2002); K. Melnikov
and A. Vainshtein, Phys. Rev. {\bf D70}, 113006 (2004).  We use
the recommended value in Ref.~\cite{davmar}.


\bibitem{ring} G.T. Danby, et al., Nucl. Instr. and Methods Phys. Res.
{\bf A 457}, 151-174 (2001).

\bibitem{quads} Y.K. Semertzidis, Nucl. Instrum. Methods Phys. Res.
{\bf A503} 458-484 (2003).

\bibitem{kicker} E. Efstathiadis, et al.,
Nucl. Inst. and Methods Phys. Res. {\bf A496} ,8-25 (2002).

\bibitem{nmr} R. Prigl, et al., Nucl. Inst. Methods Phys. Res.
{\bf A374} 118 (1996); X. Fei, V. Hughes and R. Prigl, Nucl. Inst.
Methods Phys. Res. {\bf A394}, 349 (1997).

\bibitem{liu} W.~Liu et al., Phys. Rev. Lett. {\bf 82}, 711 (1999).

\bibitem{E969} {\it A $(g-2)_{\mu}$ Experiment to $\pm0.2$ ppm
Precision}
 R.M. Carey, A. Gafarov,  I. Logashenko, K.R. Lynch,
 J.P. Miller,
  B.L. Roberts (co-spokesperson),
G. Bunce, W. Meng, W.M. Morse (resident spokesperson), Y.K.
Semertzidis, D. Grigoriev, B.I. Khazin, S.I. Redin, Yuri M.
Shatunov, E. Solodov, Y. Orlov, P. Debevec, D.W. Hertzog
(co-spokesperson), P. Kammel, R. McNabb, F. M\"ulhauser, K.L.
Giovanetti, K.P. Jungmann, C.J.G. Onderwater, S. Dhamija, T.P.
Gorringe, W. Korsch, F.E. Gray, B. Lauss, E.P. Sichtermann, P.
Cushman, T. Qian, P. Shagin, S. Dhawan and F.J.M. Farley.

\bibitem{inflector}  F. Krienen, D. Loomba and W. Meng,
Nucl. Inst. and Meth. {\bf A 283}, 5 (1989); A. Yamamoto, et al.,
Nucl. Instrum. and Methods Phys. Res. {\bf A491} 23-40 (2002).

\bibitem{detectors} S.A. Sedykh et al., Nucl. Inst. Methods Phys. Res.
 {\bf A455} 346 (2000).






\end{thebibliography}
\end{document}